# Stochastic theory of quantum mechanics and the Schrödinger equation.

Maurice GODART,
Email = maurice.godart@skynet.be

## 0. Introduction

Since the early days of quantum mechanics there were many attempts to explain the quantum phenomena within the framework of classical mechanics. One approach was foreshadowed (in 1952) by Fenyes who was struck by the analogy between quantum mechanics and the theory of Brownian motion. He claimed that particles are actually hidden beneath the orthodox Copenhagen interpretation instead of wavicles and assumed that they move not along the familiar trajectories of Newtonian mechanics, but rather along the intrinsically random sample functions ([1]) of a Markov process. Fenyes theory was not without difficulties and the reaction of the physics establishment was a few misdirected attacks, then silence. Nelson announced later (in 1966) a derivation of quantum mechanics from classical mechanics and Brownian motion. He had rediscovered Fenyes idea and rendered it more plausible.

To produce proof to his thesis Nelson first writes any wave function $\psi(\bar{x}, t)$ solution of the Schrödinger equation in the form:

$$\psi(\bar{x},t) = \exp\left[R(\bar{x},t) + iS(\bar{x},t)\right] \tag{0-1}$$

then defines two vectors $\bar{u}(\bar{x},t)$ and $\bar{v}(\bar{x},t)$ by the relations:

$$\begin{cases} m\,\bar{u}(x,t) = h/2\pi \,\overline{\text{grad}}\, R(x,t) \\ m\,\bar{v}(x,t) = h/2\pi \,\overline{\text{grad}}\, S(x,t) - e\bar{A}(x,t) \end{cases} \tag{0-2}$$

remarks that the vectors $\bar{v}_\pm(\bar{x},t) = \bar{v}(\bar{x},t) \pm \bar{u}(\bar{x},t)$ obey classical equations of a Markov process and finally claims that they coincide with the so-called forward and backward drift vectors characteristic of such a stochastic process. Then the particles are reintroduced with their trajectories identified with the sample functions of this process.

With students and co-workers the theory has been developed to the point where it covers the same domain as non-relativistic quantum mechanics. Unfortunately life is rarely so simple. Several authors have rightly raised objections against this proposed identification. The bad point is that these vectors depend on the chosen wave function and are thus not univocally determined. Worse these vectors are not well defined at the points where the wave function ψ is equal to zero. So the arguments in favour of this identification will certainly be difficult, if not impossible to swallow from the physical point of view.

From the mathematical point of view we can say that those objections originate in an erroneous use of the theory of Markov processes. We are generally not asked, except in the case of quizzes, to deduce the terms of a problem from its solution. The theory of the Markov process supposes that the drift vectors and diffusion tensor are determined univocally in agreement with reasonable principles and hypotheses. They can then be used for writing down the Fokker-Planck equation governing the probability density $p(\bar{x},t)$ and the Kolmogorov equations governing the conditional probability density $p(\bar{x}_1,t_1|\bar{x}_0,s_0)$. The final phase is to propose an acceptable physical interpretation of those densities. So we have presented somewhere else arguments supporting the idea that the conditional probability density is a better tool than the wave function for describing and explaining the evolution of microphysical systems.

---

[1] All those mysterious objects in use in the theory of stochastic processes will be introduced later.

We shall illustrate this method by walking again the path indicated by Nelson, but in the reverse direction. We shall finally show that the stochastic theory of quantum mechanics presented here is able to recover formally the solutions of the Schrödinger equation in a great number of particular cases.

## 1. Principles

We shall present the principles and main equations of the stochastic theory of quantum mechanics as prescriptions for those who would be afflicted by the rigorous study of it. We shall however start by reminding the average reader of some basic concepts of the theory of probability.

### 1.1 Probability concepts

The axiomatic definition of probability supposes that we are given an abstract space $\Omega$ (the sample space) of elements $\omega$ (the samples) endowed with a measure $dP$. A (measurable) subset E of $\Omega$ will be called an event and its probability is given by:

$$P(E) = \int_E dP \qquad (1.1\text{-}1)$$

where the integral represents the measure of the subset E. A probability is thus a particular case of a measure so that all theorems on measures can be applied immediately to it.

A real function $x(\omega)$ whose domain of definition is the sample space $\Omega$ is called a random variable if the subset:

$$E = \{\omega | a \leq x(\omega) \leq b\} \subset \Omega \qquad (1.1\text{-}2)$$

including all and only those samples $\omega$ for which we have $a \leq x(\omega) < b$ is an event for any values of a and b. The probability $P\{a \leq x(\omega) < b\}$ of this event is given by:

$$P\{a \leq x(\omega) < b\} = \int_E dP \qquad (1.1\text{-}3)$$

We shall not work within this abstract context and we shall exclusively consider the case where a probability density $p(x)$ exists that allows us to write more simply:

$$P\{a \leq x < b\} = \int_a^b p(x)\,dx \qquad (1.1\text{-}4)$$

The case of two random variables $x_1(\omega)$ and $x_2(\omega)$ can be treated in a similar way. We shall also say that the random variables $x_1(\omega)$ and $x_2(\omega)$ have a probability density $p(x_1, x_2)$ if we can write:

$$P\{a_1 \leq x_1 < b_1, a_2 \leq x_2 < b_2\} = \int_{a_1}^{b_1} \int_{a_n}^{b_n} p(x_1, x_2) dx_1 dx_2 \qquad (1.1\text{-}5)$$

The generalization of those definitions to the case of random vectors is straightforward.

We shall call stochastic process any phenomenon that evolves along the time under control of laws expressed in terms of probabilities. From the point of view of mathematics a stochastic process is a function $x(t, \omega)$ that depends on the two variables $\omega$ and t whose domains of variation are respectively the sample space $\Omega$ and a set T of real numbers and that defines a random variable $x(t_0, \omega)$ every times that a given value $t_0$ is assigned to the variable t. Conversely the function of t obtained by assigning a given value $\omega_0$ to the variable $\omega$ is called a sample function of the process. It is physically interpreted as describing the evolution of a specific random variable.

Given an arbitrary finite number of values $t_1, \cdots, t_n$ for the parameter t, we shall suppose that the corresponding random variables $x(t_1, \omega), \cdots, x(t_n, \omega)$ have a probability density that we shall denote by $p(x_1, t_1; \cdots; x_n, t_n)$. Let us immediately remark that the variables $x_1, \cdots, x_n$ are space variables with respect to which we must integrate to calculate the probabilities associated with the corresponding random variables. The simpler notation $p(x_1, \cdots, x_n)$ used previously in the case of several random variables is not recommended here because it leads very quickly to unmanageable difficulties.

## 1.2 First principle

The first principle states that the trajectories of particles are the sample functions of a Markov stochastic process. Developing this hypothesis will bring into play the probability density $p(\bar{x}, t)$ of finding the particle at the point $\bar{x}$ at time t and the conditional probability density $p(\bar{x}_1, t_1 | \bar{x}_0, t_0)$ of finding the particle at the point $\bar{x}_1$ at time $t_1$ under the condition that it was at the point $\bar{x}_0$ at time $t_0$. The probability density $p(\bar{x}_1, t_1; \bar{x}_0, t_0)$ of finding the particle at the point $\bar{x}_0$ at time $t_0$ and of finding it at the point $\bar{x}_1$ at time $t_1$ plays an important role as seen in the relations ([1]):

---

[1] By using the same symbol $p(*)$ for designating different functions that actually depend on different numbers of different arguments we indulge in an ambiguous notation deplorable, but sanctioned by custom

$$\begin{cases} p(\overline{x}_0, t_0) = \int p(\overline{x}_1, t_1; \overline{x}_0, t_0) d\overline{x}_1 \\ p(\overline{x}_1, t_1 | \overline{x}_0, t_0) = p(\overline{x}_1, t_1; \overline{x}_0, t_0) / p(\overline{x}_0, t_0) \end{cases} \qquad (1.2\text{-}1)$$

The normal and conditional probability densities are used in the definitions of the normal and conditional mathematical expectations $E\{f\}$ and $E\{f|x_0\}$ of a function $f(x,t)$ by writing:

$$\begin{cases} E\{f\} = \int f(\overline{x},t) p(\overline{x},t) d\overline{x} \\ E\{f|x_0\} = \int f(\overline{x},t) p(\overline{x},t|\overline{x}_0,t_0) d\overline{x} \end{cases} \qquad (1.2\text{-}2)$$

The Chapman-Kolmogorov equation:

$$p(\overline{x}_2, t_2 | \overline{x}_0, t_0) = \int_{-\infty}^{\infty} p(\overline{x}_2, t_2 | \overline{x}_1, t_1) p(\overline{x}_1, t_1 | \overline{x}_0, t_0) d\overline{x}_1 \qquad (1.2\text{-}3)$$

where we have either $t_0 < t_1 < t_2$ or $t_0 > t_1 > t_2$ will be considered as the defining property of a Markov process.

A Markov process is reversible in the sense that its evolution can be described not only by stochastic differential equations whose solutions build progressively along increasing times but also by stochastic differential equations whose solutions build progressively along decreasing times. This is why we shall define two drift vectors $v_\pm^i$ and two diffusion tensor $w_\pm^{ij}$ by the limits:

$$\begin{cases} \lim_{t_1 - t_0 = \pm 0} \int_{-\infty}^{\infty} (x_1^i - x_0^i) p(\overline{x}_1, t_1 | \overline{x}_0, t_0) dx_1 = v_\pm^i(\overline{x}_0, t_0)(t_1 - t_0) \\ \lim_{t_1 - t_0 = \pm 0} \int_{-\infty}^{\infty} (x_1^i - x_0^i)(x_1^j - x_0^j) p(\overline{x}_1, t_1 | \overline{x}_0, t_0) dx_1 = 2 w_\pm^{ij}(\overline{x}_0, t_0) |t_1 - t_0| \end{cases} \qquad (1.2\text{-}4)$$

where the symbol + indicates that the limits are computed under the condition $t_1 > t_0$ and where the symbol - indicates that the limits are computed under the condition $t_1 < t_0$.

These drift vectors and diffusion tensors appear as coefficients in three important differential equations. So the probability density $p(\overline{x},t)$ of a Markov process is a solution of the two versions of the Fokker-Planck equation:

$$\frac{\partial p(\overline{x},t)}{\partial t} + \sum_i \frac{\partial \left[ v_\pm^i(\overline{x}(t),t) p(\overline{x},t) \right]}{\partial x^i}$$
$$\mp \sum_i \sum_j \frac{\partial^2 \left[ w_\pm^{ij}(\overline{x}(t),t) p(\overline{x},t) \right]}{\partial x^i \partial x^j} = 0 \qquad (1.2\text{-}5)$$

while its conditional probability density $p(\overline{x}_1, t_1 | \overline{x}_0, t_0)$ is a solution of the two versions of the first Kolmogorov equation:

$$\frac{\partial p(\bar{x}_1,t_1\mid \bar{x}_0,t_0)}{\partial t_0} + \sum_i v^i_\pm(\bar{x}(t_0),t_0)\frac{\partial p(\bar{x}_1,t_1\mid \bar{x}_0,t_0)}{\partial x^i_0}$$
$$\pm \sum_i\sum_j w^{ij}_\pm(x(t_0),t_0)\frac{\partial^2 p(\bar{x}_1,t_1\mid \bar{x}_0,t_0)}{\partial x^i_0\,\partial x^j_0} = 0 \tag{1.2-6}$$

and of the two versions of the second Kolmogorov equation:

$$\frac{\partial p(\bar{x}_1,t_1\mid \bar{x}_0,t_0)}{\partial t_1} + \sum_i \frac{\partial\left[v^i_\pm(\bar{x}(t_1),t_1)p(\bar{x}_1,t_1\mid \bar{x}_0,t_0)\right]}{\partial x^i_1}$$
$$\mp \sum_i\sum_j \frac{\partial^2\left[w^{ij}_\pm(\bar{x}(t_1),t_1)p(\bar{x}_1,t_1\mid \bar{x}_0,t_0)\right]}{\partial x^i_1\,\partial x^j_1} = 0 \tag{1.2-7}$$

Let us remark that the equations involving the symbols $v_+$ and $w_+$ are related to the forward case where we have $t_1 > t_0$ while the equations involving the symbols $v_-$ and $w_-$ are related to the backward case where we have $t_1 < t_0$.

Let us finally remark that the Chapman-Kolmogorov equation (1.2-3) can be generalized to the case where either $t_1 = t_0$ or $t_2 = t_1$ provided that we impose:

$$p(\bar{x}_1,t_0\mid \bar{x}_0,t_0) = \delta(\bar{x}_1 - \bar{x}_0) \tag{1.2-8}$$

where $\delta$ is the Dirac function. The first Kolmogorov equation can then be considered as a partial differential equation of parabolic type in the variables $\bar{x}_0$ and $t_0$ for the unknown function $p(\bar{x}_1,t_1\mid \bar{x}_0,t_0)$ with the initial condition (1.2-8). In the same way the second Kolmogorov equation can be considered as a partial differential equation of parabolic type in the variables $\bar{x}_1$ and $t_1$ for the unknown function $p(\bar{x}_1,t_1\mid \bar{x}_0,t_0)$ with the same initial condition (1.2-8).

Quite unexpectedly the drift vectors $v^i_\pm$ and the diffusion tensors $w^{ij}_\pm$ are not independent. A theorem states indeed that we must have:

$$\begin{cases} w^{ij}_+(x,t) = w^{ij}_-(x,t) = w^{ij}(x,t) \\ p(x,t)u^i(x,t) = \sum_j \frac{\partial\left[p(x,t)w^{ij}(x,t)\right]}{\partial x^j} \end{cases} \tag{1.2-9}$$

where we have denoted by $w^{ij}$ the common value of $w^{ij}_+$ and $w^{ij}_-$ and where the auxiliary drift vectors $u^i$ and $v^i$ are defined by the relations

$$v^i_\pm = v^i \pm u^i \tag{1.2-10}$$

As an immediate consequence the addition members to members of the two forms of the Fokker-Planck equation leads to:

$$\frac{\partial p(\bar{x},t)}{\partial t} + \sum_i \frac{\partial\left[p(\bar{x},t)v^i(\bar{x},t)\right]}{\partial x^i} = 0 \tag{1.2-11}$$

This is obviously a continuity equation and because we have yet decided to call $p(\bar{x},t)$ a probability density it is natural to call $p(\bar{x},t)\bar{v}(\bar{x},t)$ the corresponding probability current density.

At first sight this principle appears to be nothing else that an example of circular definitions because it states on the one hand that the drift vectors and diffusion tensors can be defined only if the conditional probability density is already completely determined and states on the other hand that the conditional probability density can be defined as the solution of the Kolmogorov equations that can be written down only if the drift vectors and diffusion tensors are already perfectly determined. The next principles will help us getting out of this vicious circle.

**1.3    Second principle**

In the non-relativistic stochastic theory of quantum mechanics the diffusion tensor is given by:

$$w^{ij} = \kappa g^{ij} \tag{1.3-1}$$

with:

$$\kappa = h/4\pi m \tag{1.3-2}$$

where h is the Planck constant, where m is the mass of the particle and where $g^{ij}$ is the metric tensor. Note that we shall henceforth use exclusively Cartesian systems of co-ordinates where this metric tensor is equal to:

$$g^{ij} = \begin{cases} 1 & \text{if } i=j \\ 0 & \text{if } i \neq j \end{cases} \tag{1.3-3}$$

because it is simpler to present this principle in this context. However when we must solve concrete problems we can work with any system of generalized co-ordinates: polar, parabolic or many others yet if they are proving more convenient. The price to pay includes the replacement of all partial derivatives with respect to the Cartesian co-ordinates by covariant derivatives with respect to the generalized co-ordinates. We anticipate this conversion by adopting the notations and rules of the tensor calculus even in the case of the Cartesian co-ordinates where there is no difference between covariance and contravariance. In particular, we shall use the covariant metric tensor $g_{ij}$ to lower contravariant indices and we shall also use the contravariant metric tensor $g^{ij}$ to raise covariant indices in agreement with the formulae:

$$\begin{cases} v_i = \sum_j g_{ij} v^j \\ v^i = \sum_j g^{ij} v_j \end{cases} \tag{1.3-4}$$

It is now possible to eliminate the probability density p from the relations (1.2-9) and from the continuity equation (1.2-11). We arrive after elementary but tedious mathematical manipulations at the relation:

$$\frac{\partial u_i}{\partial t} = -\kappa \frac{\partial}{\partial x^i} \sum_j \frac{\partial v^j}{\partial x^j} - \frac{\partial}{\partial x^i} \sum_j u_j v^j \tag{1.3-5}$$

called the first Nelson equation.

## 1.4 Third principle

Stochastic mechanics will rely on a stochastic variational principle according to which the evolution of a system starting at point $\bar{x}_0$ at time $t_0$ and arriving at point $\bar{x}_1$ at time $t_1$ is described by the Markov process that makes zero the variation of the functional:

$$J(t_0, t_1) = E\left\{\int_{t_0}^{t_1} L\, dt\right\} \quad (1.4\text{-}1)$$

when the variations of the variables $x^k$ are arbitrary except that they must satisfy the boundary conditions:

$$\begin{cases} x^k(t_0) = x_0^k \\ x^k(t_1) = x_1^k \end{cases} \quad (1.4\text{-}2)$$

The functional J will act on a Lagrangian function:

$$L = L\left[x^k(t), v^k(t), u^k(t), t\right] \quad (1.4\text{-}3)$$

depending on the stochastic process $x^k(t)$, on its drift vectors $v^k(t)$ and $u^k(t)$ and on the independent variable t. The solution of this variational problem is given not by a specific sample function as we could expect from the classical theory, but by (the sample functions of) a Markov process whose drift vectors and diffusion tensor verify the Euler-Lagrange equations:

$$\frac{\partial}{\partial t}\frac{\partial L}{\partial v^k} + \sum_i v^i \frac{\partial}{\partial x^i}\frac{\partial L}{\partial v^k} - \sum_i u^i \frac{\partial}{\partial x^i}\frac{\partial L}{\partial u^k} - \sum_i \sum_j w^{ij} \frac{\partial^2}{\partial x^i \partial x^j}\frac{\partial L}{\partial u^k} = \frac{\partial L}{\partial x^k} \quad (1.4\text{-}4)$$

We shall now define the action function of the system as being equal to the mathematical expectation:

$$W(x_0, t_0; x, t) = E\left\{\int_{t_0}^{t} L(\bar{x}, \bar{v}_+, \bar{v}_-, t)\, dt\right\} \quad (1.4\text{-}5)$$

evaluated between the fixed starting point $\bar{x}_0$ and time $t_0$ and the variable ending point $\bar{x}$ and time t along sample functions of the Markov process that verifies the Euler-Lagrange equations. Calculating the derivatives of this function $W(\bar{x}_0, t_0; \bar{x}, t)$ with respect to the variables $x^i$ and with respect to the time t leads to:

$$\frac{\partial W}{\partial x^k} = \frac{\partial L}{\partial v^k} \quad (1.4\text{-}6)$$

that clearly shows that the right-hand side member must be a gradient and also to:

$$\frac{\partial W}{\partial t} = L - \sum_k v^k \frac{\partial L}{\partial v^k} - \sum_k u^k \frac{\partial L}{\partial u^k} \quad (1.4\text{-}7)$$

If applied to a physical problem the stochastic theory will simply paraphrase what is done in classical mechanics by choosing:

$$L = \frac{m}{2} \sum_i \sum_j g_{ij} \left( v^i v^j + u^i u^j \right) + e \sum_i A_i v^i - eV \qquad (1.4\text{-}8)$$

where m is (again) the mass of the particle, where e is its electrical charge and where V and $\bar{A}$ are respectively the electric and magnetic potentials acting on the particle. In the present case we have:

$$\begin{cases} \dfrac{\partial L}{\partial u^k} = m u_k \\ \dfrac{\partial L}{\partial v^k} = m v_k + e A_k \end{cases} \qquad (1.4\text{-}9)$$

and the Euler-Lagrange equations become:

$$m \left( \frac{\partial v^k}{\partial t} + \sum_i v^i \frac{\partial v^k}{\partial x^i} - \sum_i u^i \frac{\partial u^k}{\partial x^i} - \sum_i \sum_j w^{ij} \frac{\partial^2 u^k}{\partial x^i \partial x^j} \right)$$
$$= e E_k + e \sum_i B_{ik} v^i \qquad (1.4\text{-}10)$$

where:

$$E_k = -\frac{\partial V}{\partial x^k} - \frac{\partial A_k}{\partial t} \qquad (1.4\text{-}11)$$

is the electric field and where

$$B_{ik} = \frac{\partial A_i}{\partial x^k} - \frac{\partial A_k}{\partial x^i} \qquad (1.4\text{-}12)$$

is the magnetic field. The equation (1.4-10) is called the second Nelson equation. Let us remark that its right hand side member represents the Lorentz force acting on the particle.

Note in passing that as it results from the relations (1.4-6) and (1.4-9) that we have:

$$m v_k + e A_k = \frac{\partial W}{\partial x^k} \qquad (1.4\text{-}13)$$

showing that $m\bar{v} + e\bar{A}$ must be a gradient.

The gradual transition (conceptual and practical) from the microscopic level with its quantum laws to the macroscopic level with its classical laws suggests that any quantum theory must reproduce the classical theory at some suitable approximation. The criterion proposed by the orthodox theory is limit $h = 0$ while the one proposed by the stochastic theory is the more realistic limit $m = \infty$. Both lead to the limit $w^{ij} = 0$. We can then write:

$$\begin{cases} u^i = 0 \\ w^{ij} = 0 \\ v^i_\pm = v^i = dx^i / dt \end{cases} \qquad (1.4\text{-}14)$$

and the Euler-Lagrange equations take the well-known classical form:

$$m\frac{dv^k}{dt} = eE_k + e\sum_i B_{ik}v^i \tag{1.4-15}$$

## 2. The way to Schrödinger equation

We shall now tackle the problem suggested by the title of this publication, namely that we shall determine the hypotheses with which it is possible to derive the Schrödinger equation from the stochastic theory of quantum mechanics.

### 2.1 Stationary Markov processes.

We shall examine here the properties of the solutions of the Kolmogorov and Fokker-Planck equations in the particular case where the drift vectors $v^i_\pm(x)$ and the diffusion tensor $w^{ij}(x)$ do not actually depend on the time. Then the equation (1.2-9) can be used to prove that the probability density $p(x)$ also does not actually depend on the time.

We shall restrict our study to the case of the forward equations because all that will be said about them is also valid for the backward equations with due changes. Thus taking into account the preliminary remarks, we see that the Fokker-Planck equation governing the probability density $p(\bar{x})$ reduces to:

$$\sum_i \frac{\partial \left[ v^i_+(\bar{x})p(\bar{x}) \right]}{\partial x^i} - \sum_i \sum_j \frac{\partial^2 \left[ w^{ij}(\bar{x})p(\bar{x}) \right]}{\partial x^i \partial x^j} = 0 \tag{2.1-1}$$

and that the Kolmogorov equations governing the conditional probability density $p(\bar{x}_1, t_1 \mid \bar{x}_0, t_0)$ are:

$$\frac{\partial p(\bar{x}_1, t_1 \mid \bar{x}_0, t_0)}{\partial t_0} + \sum_i v^i_+(\bar{x}_0) \frac{\partial p(\bar{x}_1, t_1 \mid \bar{x}_0, t_0)}{\partial x^i_0}$$
$$+ \sum_i \sum_j w^{ij}(\bar{x}_0) \frac{\partial^2 p(\bar{x}_1, t_1 \mid \bar{x}_0, t_0)}{\partial x^i_0 \partial x^j_0} = 0 \tag{2.1-2}$$

$$\frac{\partial p(\bar{x}_1, t_1 \mid \bar{x}_0, t_0)}{\partial t_1} + \sum_i \frac{\partial \left[ v^i_+(\bar{x}_1) p(\bar{x}_1, t_1 \mid \bar{x}_0, t_0) \right]}{\partial x^i_1}$$
$$- \sum_i \sum_j \frac{\partial^2 \left[ w^{ij}(\bar{x}_1) p(\bar{x}_1, t_1 \mid \bar{x}_0, t_0) \right]}{\partial x^i_1 \partial x^j_1} = 0 \tag{2.1-3}$$

Those equations can be solved by the method of separation of the variables. Remember that this amounts to represent their general solution as a linear combination of particular solutions that are products $X^0(\bar{x}_0)X^1(\bar{x}_1)T^0(t_0)T^1(t_1)$ of functions depending on a single variable at a time. If we introduce such a product into the equations, divide immediately by this same product and simplify the obtained expression we arrive at the equivalent equations:

$$\begin{cases} -\dfrac{\dot{T}^0(t_0)}{T^0(t_0)} = \dfrac{1}{X^0(\overline{x}_0)}\left[\sum_i\sum_j w^{ij}(\overline{x}_0)\dfrac{\partial^2 X^0(\overline{x}_0)}{\partial x_0^i\,\partial x_0^j} + \sum_i v_+^i(\overline{x}_0)\dfrac{\partial X^0(\overline{x}_0)}{\partial x_0^i}\right] \\[2ex] \dfrac{\dot{T}^1(t_1)}{T^1(t_1)} = \dfrac{1}{X^1(\overline{x}_1)}\left[\sum_i\sum_j \dfrac{\partial^2\left[w^{ij}(\overline{x}_1)X^1(\overline{x}_1)\right]}{\partial x_1^i\,\partial x_1^j} + \sum_i \dfrac{\partial\left[v_+^i(\overline{x}_1)X^1(\overline{x}_1)\right]}{\partial x_1^i}\right] \end{cases} \quad (2.1\text{-}4)$$

where the dot designates the derivation with respect to the variables $t_0$ or $t_1$. The left hand side members of those equations do not depend on the variable $\overline{x}$ while their right hand side member does not depend on the variable t so that their common value must be a constant that we shall designate respectively by $\lambda$ and $\mu$. Then the equations for $T^0(t_0)$ and $T^1(t_1)$ can be solved immediately and lead to:

$$\begin{cases} T^0(t_0) = \exp(-\lambda t_0) \\ T^1(t_1) = \exp(\mu t_1) \end{cases} \quad (2.1\text{-}5)$$

while the equations for $X^0(\overline{x}_0)$ and $X^1(\overline{x}_1)$ take the forms:

$$\begin{cases} \sum_i\sum_j w^{ij}(\overline{x}_0)\dfrac{\partial^2 X^0(\overline{x}_0)}{\partial x_0^i\,\partial x_0^j} + \sum_i v_+^i(\overline{x}_0)\dfrac{\partial X^0(\overline{x}_0)}{\partial x_0^i} = \lambda X^0(\overline{x}_0) \\[2ex] \sum_i\sum_j \dfrac{\partial^2\left[w^{ij}(\overline{x}_1)X^1(\overline{x}_1)\right]}{\partial x_1^i\,\partial x_1^j} - \sum_i \dfrac{\partial\left[v_+^i(\overline{x}_1)X^1(\overline{x}_1)\right]}{\partial x_1^i} = \mu X^1(\overline{x}_1) \end{cases} \quad (2.1\text{-}6)$$

**2.2  Theoretical interlude**

Before proceeding any further we think it is a good idea to review the basic properties of the second order linear differential operator:

$$L(y) = \sum_i\sum_j a^{ij}(\overline{x})\dfrac{\partial^2 y(\overline{x})}{\partial x^i\partial x^j} + \sum_i a^i(\overline{x})\dfrac{\partial y(\overline{x})}{\partial x^i} + a(\overline{x})y(\overline{x}) \quad (2.2\text{-}1)$$

The corresponding adjoint differential operator $M(z)$ is defined by the condition that the so-called Lagrange identity:

$$zL(y) - yM(z) = \sum_i \dfrac{\partial P^i(y,z)}{\partial x^i} \quad (2.2\text{-}2)$$

is verified where the functions $P^i(y,z)$ are bilinear forms in the functions y and z and their first order partial derivatives. The operator $M(z)$ and the functions $P^i(y,z)$ so defined are univocally determined simply by integrating the expression $zL(y)$ by parts in order to remove all its partial derivatives of y. Elementary mathematical manipulations show that the relation (2.2-2) is satisfied provided we choose:

$$P^i(y,z) = \sum_j a^{ij}\left(z\dfrac{\partial y}{\partial x^j} - y\dfrac{\partial z}{\partial x^j}\right) + \left(a^i - \sum_j \dfrac{\partial}{\partial x^j}a^{ij}\right)yz \quad (2.2\text{-}3)$$

and:

$$M(y) = \sum_i \sum_j \frac{\partial^2 a^{ij}(\bar{x}) y(\bar{x})}{\partial x^i \partial x^j} - \sum_i \frac{\partial a^i(\bar{x}) y(\bar{x})}{\partial x^i} + a(\bar{x}) y(\bar{x}) \qquad (2.2\text{-}4)$$

We can also write it in the form:

$$M(y) = \sum_i \sum_j b^{ij}(\bar{x}) \frac{\partial^2 y(\bar{x})}{\partial x^i \partial x^j} + \sum_i b^i(\bar{x}) \frac{\partial y(\bar{x})}{\partial x^i} + b(\bar{x}) y(\bar{x}) \qquad (2.2\text{-}5)$$

with:

$$\begin{cases} b^{ij}(\bar{x}) = a^{ij}(\bar{x}) \\ b^i(\bar{x}) = 2 \sum_j \frac{\partial a^{ij}(\bar{x})}{\partial x^j} - a^i(\bar{x}) \\ b(\bar{x}) = \sum_i \sum_j \frac{\partial^2 a^{ij}(\bar{x})}{\partial x^i \partial x^j} - \sum_i \frac{\partial a^i(\bar{x})}{\partial x^i} + a(\bar{x}) \end{cases} \qquad (2.2\text{-}6)$$

In view of the symmetry between the operators L and M in the Lagrange identity (2.2-2) it is obvious that they are adjoint to each other.

We shall suppose that the quadratic form built with the coefficients $a^{ij}(\bar{x})$ is positive definite so that the operators $L(y)$ and $M(z)$ are elliptic. Let us now consider the two equations:

$$L[y(\bar{x})] = \lambda y(\bar{x}) \qquad (2.2\text{-}7)$$

and

$$M[z(\bar{x})] = \mu z(\bar{x}) \qquad (2.2\text{-}8)$$

where $\lambda$ and $\mu$ are constants. When those equations are supplemented with the condition:

$$\int [zL(y) - yM(z)] d\bar{x} = 0 \qquad (2.2\text{-}9)$$

we can claim that solutions exist and exhibit the following properties.

Firstly, the equation (2.2-7) has non-identically zero solutions $y_i(\bar{x})$ called eigenfunctions only for a discrete countably infinite set of eigenvalues $\lambda_i$. Similarly, the equation (2.2-8) has non-identically zero solutions $z_j(\bar{x})$ called eigenfunctions only for a discrete countably infinite set of eigenvalues $\mu_j$.

Secondly we can claim that any particular eigenvalue $\lambda_i$ associated with the eigenfunction $y_i(\bar{x})$ of the equation (2.2-7) must be equal to one of the eigenvalues $\mu_j$ associated with the eigenfunction $z_j(\bar{x})$ of the equation (2.2-8) and that conversely any particular eigenvalue $\mu_j$ associated with the eigenfunction $z_j(\bar{x})$ of the equation (2.2-8) must be equal to one of the eigenvalues $\lambda_i$ associated with the eigenfunction $y_i(\bar{x})$ of the equation (2.2-7).

Thirdly this correspondence between the eigenvalues $\lambda_i$ and the eigenvalues $\mu_j$ allows us to suppose that the indices for those are the non negative integers chosen so that the sets of the eigenvalues $(\lambda_i)$ and $(\mu_j)$ verify the equalities:

$$\lambda_i = \mu_i \tag{2.2-10}$$

This correspondence can be extended to the associated eigenfunctions $y_i(\overline{x})$ and $z_j(\overline{x})$. Remarking that they are defined except for constant factors they can be normalized so that they verify the orthonormalization conditions:

$$\int y_i(\overline{x}) z_j(\overline{x}) d\overline{x} = \delta_{i\,j} \tag{2.2-11}$$

with:

$$\delta_{ij} = \begin{cases} 0 & \text{if } i \neq j \\ 1 & \text{if } i = j \end{cases} \tag{2.2-12}$$

The operators $L(y)$ and $M(z)$ are said to be self-adjoint if we have identically:

$$M(y) = L(y) \tag{2.2-13}$$

for any function $y(x)$ and according to the relations (2.2-6), the necessary and sufficient conditions for this identification to be possible is simply:

$$a^i(\overline{x}) = \sum_j \frac{\partial a^{ij}(\overline{x})}{\partial x^j} \tag{2.2-14}$$

The elliptic self-adjoint operators have been studied extensively and their theory states that all their eigenvalues and eigenfunctions are real. This is a welcome property and the question naturally arises to know if it can be extended to the case of non-self-adjoint operators. The answer is generally in the negative with the exception however of these operators $L(y)$ for which it is possible to find two functions $\alpha(x)$ and $\beta(x)$ such that the modified operator defined by:

$$\tilde{L}(z) = \alpha L(\beta z) \tag{2.2-15}$$

is self-adjoint. If we introduce the expression $y(\overline{x}) = \beta(\overline{x}) z(\overline{x})$ in the operator $L(y)$ we obtain after some elementary manipulations:

$$\tilde{L}(z) = \sum_i \sum_j c^{ij} \frac{\partial^2 z}{\partial x^i \partial x^j} + \sum_i a^i \frac{\partial z}{\partial x^i} + c\,z \tag{2.2-16}$$

with

$$\begin{cases} c^{ij} = \alpha \beta a^{ij} \\ c^i = \alpha \left( 2 \sum_j a^{ij} \frac{\partial \beta}{\partial x^j} + a^i \beta \right) \\ c = \alpha \left( \sum_i \sum_j a^{ij} \frac{\partial^2 \beta}{\partial x^i \partial x^j} + \sum_i a^i \frac{\partial \beta}{\partial x^i} + a \beta \right) \end{cases} \tag{2.2-17}$$

Applying the general criteria (2.2-14) to this operator shows that the necessary and sufficient conditions for it to be self-adjoint is:

$$\sum_j a^{ij} \frac{\partial}{\partial x^j} \ln \frac{\alpha}{\beta} = a^i - \sum_j \frac{\partial a^{ij}}{\partial x^j} \quad (2.2\text{-}18)$$

Let us remark that only the ratio $\alpha/\beta$ is relevant, so that the existence of one solution does actually imply the existence of many others and that we can impose the additional condition:

$$\alpha \beta = 1 \quad (2.2\text{-}19)$$

that by the way makes the operator $\tilde{L}(z)$ elliptic as this results from the definition (2.2-16). Quite curiously the same functions $\alpha(x)$ and $\beta(x)$ can be reused to make the modified operator:

$$\tilde{M}(z) = \beta M(\alpha z) \quad (2.2\text{-}20)$$

elliptic and self-adjoint too. We can also write the Lagrange identity in the form:

$$z\tilde{L}(y) - y\tilde{M}(z) = \alpha z L(\beta y) - \beta y M(\alpha z)$$
$$= \sum_i \frac{\partial P^i(\beta y, \alpha z)}{\partial x^i} \quad (2.2\text{-}21)$$

showing that the operators $\tilde{L}$ and $\tilde{M}$ are adjoint to each other. Thus the operators $\tilde{L}$ and $\tilde{M}$ must be identical if one of them is self-adjoint. We shall henceforth designate by $N(z)$ their common expression and thus suppose implicitly that the equation (2.2-18) possesses at least one solution.

We have proved that the operator $N(z)$ is elliptic and self-adjoint in which case the classical theory of the elliptic equations states that all its eigenvalues $\lambda_i$ and eigenfunctions $z_i(x)$ that thus satisfy the equations:

$$N(z_i) = \lambda_i z_i \quad (2.2\text{-}22)$$

are real. The same conclusion is then valid for the operator L as it results from the equalities:

$$L(\beta z_i) = \alpha \beta L(\beta z_i)$$
$$= \beta N(z_i) \quad (2.2\text{-}23)$$
$$= \lambda_i (\beta z_i)$$

showing that $\lambda_i$ is the (real) eigenvalue associated with the (real) eigenfunction $\beta z_i$ of the operator L The same conclusion is valid for the operator M as it results from the equalities:

$$M(\alpha z_i) = \lambda_i (\alpha z_i) \quad (2.2\text{-}24)$$

showing that $\lambda_i$ is also the (real) eigenvalue associated with the (real) eigenfunction $\alpha z_i$ of the operator M.

## 2.3 Stationary Markov processes (continued)

We can apply those general results to the stationary Markov processes provided that we perform the substitutions:

$$\begin{cases} a(\bar{x}) = 0 \\ a^i(\bar{x}) = v^i_+(\bar{x}) \\ a^{ij}(\bar{x}) = w^{ij}(\bar{x}) \end{cases} \quad (2.3\text{-}1)$$

All that has been said up to now can be applied to the equations (2.1-6). They are obviously adjoint to each other and elliptic because the quadratic form built with the coefficients $w^{ij}$ is positive definite.

With the additional substitutions;

$$\begin{cases} y_i(\bar{x}) = X_i^0(\bar{x}) \\ z_i(\bar{x}) = X_i^1(\bar{x}) \end{cases} \quad (2.3\text{-}2)$$

we can write:

$$\int X_i^0(\bar{x}) X_j^1(\bar{x}) d\bar{x} = \delta_{ij} \quad (2.3\text{-}3)$$

and the general solution $p(\bar{x}_1, t_1 | \bar{x}_0, t_0)$ of the Kolmogorov equations takes the form:

$$p(\bar{x}_1, t_1 | \bar{x}_0, t_0) = \sum_i \sum_j A_{ij} X_i^0(\bar{x}_0) X_j^1(\bar{x}_1) \exp(-\lambda_i t_0) \exp(\lambda_j t_1) \quad (2.3\text{-}4)$$

However because of the initial condition (1.2-8) and the orthonormalization conditions (2.3-3) we can write:

$$A_{ij} = \delta_{ij} \quad (2.3\text{-}5)$$

and we finally obtain:

$$p(\bar{x}_1, t_1 | \bar{x}_0, t_0) = \sum_i X_i^0(\bar{x}_0) X_i^1(\bar{x}_1) \exp\left[\lambda_i (t_1 - t_0)\right] \quad (2.3\text{-}6)$$

The question now arises to know when we can claim that all the eigenvalues and eigenfunctions of the equations (2.2-7) and (2.2-8) are real. In the present case the general conditions (2.2-18) can be written in the form:

$$\sum_j w^{ij}(\bar{x}) \frac{\partial}{\partial x^j} \ln(\alpha/\beta) = v^i_+(\bar{x}) - \sum_j \frac{\partial w^{ij}(\bar{x})}{\partial x^j} \quad (2.3\text{-}7)$$

It is obvious that the function $X_o(\bar{x}) = 1$ is an eigenfunction for the equation (2.2-7) and that the function $X_1(\bar{x}) = p(\bar{x})$ is an eigenfunction of the equation (2.2-8) associated both with the eigenvalue 0. Thus, if equation (2.2-18) possesses a solution, we know that an eigenfunction $z(\bar{x})$ of the operators $N(z)$ must exist such that:

$$\begin{cases} \beta z(\bar{x}) = 1 \\ \alpha z(\bar{x}) = p(\bar{x}) \end{cases} \quad (2.3\text{-}8)$$

showing that the solution must be:

$$\alpha/\beta = p(\bar{x}) \tag{2.3-9}$$

Actually this necessary condition is not sufficient. And indeed if we introduce this expression of the solution in equation (2.3-7) we easily obtain:

$$v^i_+(\bar{x}) = u^i(\bar{x}) \tag{2.3-10}$$

and according to (1.2-10) this implies that we must have:

$$v^{i\cdot}(\bar{x}) = 0 \tag{2.3-11}$$

We can conclude that this condition is necessary and sufficient for guaranteeing the existence of a solution $\alpha/\beta = p(\bar{x})$ and so for guaranteeing that the operator $N(z)$ exists.

Let us briefly mention the existence of an H-theorem proving that we must have:

$$\begin{aligned}\lim_{t_1=\infty} p(\bar{x},t \mid \bar{x}_0,t_0) &= \lim_{t_1=\infty} \sum_i X_i^1(\bar{x}_1) X_i^0(\bar{x}_0) \exp\left[\lambda_i(t_1-t_0)\right] \\ &= X_0^1(\bar{x}_1) X_0^0(\bar{x}_0) \\ &= p(\bar{x})\end{aligned} \tag{2.3-12}$$

showing clearly that the physical phenomenon described by the Kolmogorov equations has an inherent tendency to go or to return to the stable state characterized by the probability density $p(\bar{x})$. This is possible if and only if all the eigenvalues other than $\lambda = \mu = 0$ are negative.

## 2.4 Non-relativistic stationary Markov processes.

We shall study here the solutions of the Kolmogorov and Fokker-Planck equations in the particular case where the electric field $\bar{E}(\bar{x})$ and magnetic field $\bar{B}(\bar{x})$ acting on the particle do not actually depend on the time.

According to the second principle the diffusion tensor $w^{ij}(\bar{x})$ does not depend on the time. Moreover the structure of the Nelson equations suggests that they can be verified by drift vectors $v^i_\pm(\bar{x})$ that do not depend on the time either. The corresponding Markov process is thus stationary so that we can apply to it all the general results obtained in the previous paragraph. As a first result, we know that the probability density $p(\bar{x})$ does not actually depend on the time.

We shall now consider the particular case where $v_k = 0$ only in order to take advantage of all the pleasant properties of the operator $N(z)$. With the substitutions:

$$\begin{cases} a(\bar{x}) = 0 \\ a^i(\bar{x}) = u^i(\bar{x}) \\ a^{ij}(\bar{x}) = w^{ij}(\bar{x}) \end{cases} \tag{2.4-1}$$

the Fokker-Planck equation governing the probability density $p(\bar{x})$ takes the form:

$$\sum_i \frac{\partial\left[u^i(\bar{x})p(\bar{x})\right]}{\partial x^i} - \sum_i \sum_j \frac{\partial^2\left[w^{ij}(\bar{x})p(\bar{x})\right]}{\partial x^i \partial x^j} = 0 \tag{2.4-2}$$

and the Kolmogorov equations governing the conditional probability density $p(\bar{x},t\,|\,\bar{x}_0,t_0)$ take the forms:

$$\frac{\partial p(\bar{x}_1,t_1\,|\,\bar{x}_0,t_0)}{\partial t_0}+\sum_i u^i(\bar{x}_0)\frac{\partial p(\bar{x}_1,t_1\,|\,\bar{x}_0,t_0)}{\partial x_0^i}$$
$$+\sum_i\sum_j w^{ij}(\bar{x}_0)\frac{\partial^2 p(\bar{x}_1,t_1\,|\,\bar{x}_0,t_0)}{\partial x_0^i\,\partial x_0^j}=0 \qquad (2.4\text{-}3)$$

$$\frac{\partial p(\bar{x}_1,t_1\,|\,\bar{x}_0,t_0)}{\partial t_1}+\sum_i\frac{\partial\left[u^i(\bar{x}_1)p(\bar{x}_1,t_1\,|\,\bar{x}_0,t_0)\right]}{\partial x_1^i}$$
$$-\sum_i\sum_j\frac{\partial^2\left[w^{ij}(\bar{x}_1)p(\bar{x}_1,t_1\,|\,\bar{x}_0,t_0)\right]}{\partial x_1^i\,\partial x_1^j}=0 \qquad (2.4\text{-}4)$$

Those equations can be solved by the method of separation of the variables as explained before. The equations (2.1-6) take the form:

$$\begin{cases}\displaystyle\sum_j\sum_j w^{ij}(\bar{x}_0)\frac{\partial^2 X^0(\bar{x}_0)}{\partial x_0^i\,\partial x_0^j}+\sum_i u^i(\bar{x}_0)\frac{\partial X^0(\bar{x}_0)}{\partial x_0^i}+\lambda X^0(\bar{x}_0)=0\\ \displaystyle\sum_i\sum_j\frac{\partial^2\left[w^{ij}(\bar{x}_1)X^1(\bar{x}_1)\right]}{\partial x_1^i\,\partial x_1^j}-\sum_i\frac{\partial\left[u^i(\bar{x}_1)X^1(\bar{x}_1)\right]}{\partial x_1^i}+\mu X^1(\bar{x}_1)=0\end{cases} \qquad (2.4\text{-}5)$$

They exhibit many interesting properties. So for example all their eigenvalues and eigenfunctions are real. Moreover $X^0(\bar{x}_0)$ is an eigenfunction of the first equation associated with the eigenvalue $\lambda$ if and only if $X(\bar{x})=p(\bar{x})X^0(\bar{x})$ is an eigenfunction of the second equation associated with the same eigenvalue $\mu=\lambda$ We have indeed:

$$\frac{\partial^2(pXw^{ij})}{\partial x^i\partial x^j}-\frac{\partial(pXu^i)}{\partial x^i}+\lambda pX$$
$$=pw^{ij}\frac{\partial^2 X}{\partial x^i\partial x^j}+2\frac{\partial(pw^{ij})}{\partial x^j}\frac{\partial X}{\partial x^i}+\frac{\partial^2(pw^{ij})}{\partial x^i\partial x^j}X$$
$$-pu^i\frac{\partial X}{\partial x^i}-\frac{\partial(pu^i)}{\partial x^i}X+\lambda pX \qquad (2.4\text{-}6)$$
$$=p\left(w^{ij}\frac{\partial^2 X}{\partial x^i\partial x^j}+u^i\frac{\partial X}{\partial x^i}+\lambda X\right)$$

Let us now prove an all-important result. Let $X^0$ be any eigenfunction of the first equation in the system (2.4-5) associated with the eigenvalue $\lambda$. Then the function:

$$\Psi=\sqrt{p}\,X^0 \qquad (2.4\text{-}7)$$

for which we have:

$$\sqrt{p}\,\Psi=pX^0 \qquad (2.4\text{-}8)$$

obviously verifies the second equation in the system (2.4-5). From the relation (1.2-9) we can deduce that:

$$2\frac{\partial\left(w^{ij}\sqrt{p}\right)}{\partial x^{j}} = \frac{1}{\sqrt{p}}\frac{\partial\left(w^{ij}p\right)}{\partial x^{j}} \qquad (2.4\text{-}9)$$
$$= u^{i}\sqrt{p}$$

and that:

$$2\frac{\partial^{2} w^{ij}\sqrt{p}}{\partial x^{i}\partial x^{j}} = \frac{\partial u^{i}\sqrt{p}}{\partial x^{j}} \qquad (2.4\text{-}10)$$

We shall start by writing:

$$\begin{aligned}
0 &= \frac{\partial^{2}\left(w^{ij}\sqrt{p}\,\Psi\right)}{\partial x^{i}\partial x^{j}} - \frac{\partial\left(u^{i}\sqrt{p}\,\Psi\right)}{\partial x^{i}} + \lambda\sqrt{p}\,\Psi \\
&= w^{ij}\sqrt{p}\frac{\partial^{2}\Psi}{\partial x^{i}\partial x^{j}} + 2\frac{\partial\left(w^{ij}\sqrt{p}\right)}{\partial x^{j}}\frac{\partial\Psi}{\partial x^{i}} + \frac{\partial^{2}\left(w^{ij}\sqrt{p}\right)}{\partial x^{i}\partial x^{j}}\Psi \\
&\quad - u^{i}\sqrt{p}\frac{\partial\Psi}{\partial x^{i}} - \frac{\partial\left(u^{i}\sqrt{p}\right)}{\partial x^{i}}\Psi + \lambda\sqrt{p}\,\Psi \\
&= w^{ij}\sqrt{p}\frac{\partial^{2}\Psi}{\partial x^{i}\partial x^{j}} - \frac{1}{2}\frac{\partial u^{i}\sqrt{p}}{\partial x^{i}}\Psi + \lambda\sqrt{p}\,\Psi
\end{aligned} \qquad (2.4\text{-}11)$$

Using now the two relations:

$$\begin{aligned}
u_{k}\sqrt{p} &= 2g_{ik}\frac{\partial\left(w^{ij}\sqrt{p}\right)}{\partial x^{j}} \\
&= 2\kappa\delta_{k}^{j}\frac{\partial\sqrt{p}}{\partial x^{j}} \\
&= 2\kappa\frac{\partial\sqrt{p}}{\partial x^{k}}
\end{aligned} \qquad (2.4\text{-}12)$$

and:

$$\begin{aligned}
\frac{\partial\left(u^{j}\sqrt{p}\right)}{\partial x^{j}} &= \frac{\partial\sqrt{p}}{\partial x^{j}}u^{j} + \sqrt{p}\frac{\partial u^{j}}{\partial x^{j}} \\
&= \sqrt{p}\left(\frac{1}{2\kappa}u^{j}u_{j} + g^{ij}\frac{\partial u_{i}}{\partial x^{j}}\right)
\end{aligned} \qquad (2.4\text{-}13)$$

we shall continue by writing:

$$0 = w^{ij}\sqrt{p}\,\frac{\partial^2 \Psi}{\partial x^i \partial x^j} - \frac{1}{2}\frac{\partial u^i \sqrt{p}}{\partial x^i}\Psi + \lambda\sqrt{p}\,\Psi$$

$$= \sqrt{p}\left(w^{ij}\frac{\partial^2 \Psi}{\partial x^i \partial x^j} - \frac{1}{2}\frac{\partial u^i}{\partial x^i}\Psi - \frac{u^i}{2\sqrt{p}}\frac{\partial \sqrt{p}}{\partial x^i}\Psi + \lambda\Psi\right) \quad (2.4\text{-}14)$$

$$= \sqrt{p}\left[w^{ij}\frac{\partial^2 \Psi}{\partial x^i \partial x^j} - \frac{1}{2}\left(\frac{1}{2\kappa}u^i u_i + g^{ij}\frac{\partial u_j}{\partial x^i}\right)\Psi + \lambda\Psi\right]$$

We shall calculate the second term within the parentheses indirectly by evaluating its partial derivative with respect to $x^k$. Note first the relation (2.4-12) can be written in the form:

$$u_k = 2\kappa\frac{\partial \ln \sqrt{p}}{\partial x^k} \quad (2.4\text{-}15)$$

showing that $u_i$ is a gradient. We then obtain:

$$\frac{\partial}{\partial x^k}\left(\frac{1}{2\kappa}u^i u_i + g^{ij}\frac{\partial u_j}{\partial x^i}\right) = \frac{1}{\kappa}u^i\frac{\partial u_i}{\partial x^k} + g^{ij}\frac{\partial^2 u_j}{\partial x^i \partial x^k}$$
$$= \frac{1}{\kappa}u^i\frac{\partial u_k}{\partial x^i} + g^{ij}\frac{\partial^2 u_k}{\partial x^i \partial x^j} \quad (2.4\text{-}16)$$

The Euler-Lagrange equations allow us to write:

$$m\left(u^i\frac{\partial u_k}{\partial x^i} + w^{ij}\frac{\partial^2 u_k}{\partial x^i \partial x^j}\right) = e\frac{\partial V}{\partial x^k} \quad (2.4\text{-}17)$$

and we thus can write:

$$\frac{\partial}{\partial x^k}\left(\frac{1}{2\kappa}u^i u_i + g^{ij}\frac{\partial u_j}{\partial x^i}\right) = \frac{e}{\kappa m}\frac{\partial V}{\partial x^k} \quad (2.4\text{-}18)$$

or:

$$\frac{1}{2\kappa}u^i u_i + g^{ij}\frac{\partial u_j}{\partial x^i} = \frac{e}{\kappa m}V \quad (2.4\text{-}19)$$

and we so arrive at the equation:

$$0 = w^{ij}\frac{\partial^2 \Psi}{\partial x^i \partial x^j} - \frac{e}{2\kappa m}V\Psi + \lambda\Psi \quad (2.4\text{-}20)$$

Multiplying by $2m\kappa$, introducing the new constant:

$$E = 2m\kappa\lambda \quad (2.4\text{-}21)$$

and noting that we have:

$$2\kappa m w^{ij} = \frac{h^2}{8\pi^2 m} \quad (2.4\text{-}22)$$

we arrive finally at:

$$\frac{h^2}{8\pi^2 m}\Delta\Psi = (E-V)\Psi \tag{2.4-23}$$

which is nothing else than the Schrödinger equation independent on the time.

## 3. Conclusion

To be sure we did not prove and we did not intend to prove that the Schrödinger equation is equivalent to the Kolmogorov equations. Remember that the former can be deduced from the latter only if we hypothesize that the underlying Markov process is stationary and that its drift vector $\bar{v}$ is equal to zero. At most can we say that there exists a hazy relationship between their solutions.

The second mathematical condition $\bar{v}=0$ is rather abstract and looks artificial, but it can be immediately transformed into a more familiar physical condition. We have seen indeed as a result of our stochastic variational principle that the vector $m\bar{v}+e\bar{A}$ must be a gradient. Thus the magnetic potential $\bar{A}$ must also be a gradient if $\bar{v}=0$ and the magnetic field $\bar{B}=\operatorname{rot}\bar{A}$ must inevitably be equal to zero.

Conversely in the case where the magnetic field is different from zero, the velocity $\bar{v}$ must also be different from zero. In this case the equation (2.3-7) has no solution and the operator $N(z)$ does not exist. If so the non-zero eigenvalues $\lambda_i = \mu_i$ and associated eigenfunctions $X_i^0(\bar{x})$ and $X_i^1(\bar{x})$ of the equations (2.2-7) and (2.2-8) are (possibly) complex. We shall then propose to modify the notations by letting the index i run from $-\infty$ to $+\infty$ with the conditions:

$$\begin{cases} \lambda_{-i} = \tilde{\lambda}_i \\ X_{-i}^0(\bar{x}) = \tilde{X}_i^0(\bar{x}) \\ X_{-i}^1(\bar{x}) = \tilde{X}_i^1(\bar{x}) \end{cases}$$

where the symbol ~ indicates the complex conjugate of a mathematical expression. They can be associated so that they verify the orthonormalization conditions:

$$\int X_i^0(\bar{x})\tilde{X}_j^1(\bar{x})d\bar{x} = \delta_{ij} \tag{2.4-1}$$

and we must accordingly modify the expression of the solution (2.3-6)) by writing it in the form:

$$p(\bar{x}_1,t_1\mid\bar{x}_0,t_0) = \sum_i X_i^0(\bar{x}_0)X_i^1(\bar{x}_1)\exp\left[\lambda_i(t_1-t_0)\right]$$

Here also the H-theorem states that we must have:

$$\lim_{t_1=\infty} p(\bar{x},t\mid\bar{x}_0,t_0) = \lim_{t_1=\infty}\sum_n X_n^1(\bar{x}_1)X_n^0(\bar{x}_0)\exp\left[\lambda_n(t_1-t_0)\right]$$
$$= X_0^1(\bar{x}_1)X_0^0(\bar{x}_0)$$
$$= p(\bar{x})$$

with the same remarks as before. This is possible now if and only if (the real parts of) all the eigenvalues other than $\lambda=\mu=0$ are negative.

Our intuition leads us to think that the conjugate solutions must have something in common. In the case of the hydrogen atom for example, we can claim using the parlance of the orthodox theory that the (real) eigenfunctions $X_0^0(\bar{x}_0) = 1$ and $X_0^1(\bar{x}) = p(\bar{x})$ associated with the eigenvalues $\lambda_0 = 0$ corresponds to the ground-state and we can imagine that all other pairs $X_{-i}^0(\bar{x})$ and $X_i^0(\bar{x})$ as well as the other pairs $X_{-i}^1(\bar{x})$ and $X_i^1(\bar{x})$ associated with the eigenvalues $\lambda_{-i} = \tilde{\lambda}_i \neq 0$ describe states having some physical characteristics in common (like the energy?) and possessing some other physical characteristics in their own right (like the spin?). This idea is supported by the remark that we were unable to recover the Schrödinger equation precisely in the circumstance where the magnetic field is not equal to zero and when it must give way to the Pauli equation.

It would be highly desirable to check those ideas by studying the case of a constant magnetic field for example. However we did not succeed in solving this very simple problem and we must even confess that out of sheer laziness we threw up the sponge, hoping that we should be able to take up this challenge more easily (?) and more correctly in the context of the stochastic theory of relativistic quantum mechanics.

## 4. Afterwards

We must humbly acknowledge that we are indebted to Nelson of three important ideas. Firstly he was the one who revived the hypothesis that a Markov process was underlying quantum mechanics. No need to say that our version of the stochastic theory relies heavily on it. Secondly we owe entirely to him the expression of the diffusion tensor proposed in the second principle. Thirdly his method escapes all the objections mentioned in the introduction when it is applied only to the ground-state wave function and it then leads to vectors $\bar{u}$ and $\bar{v}$ that are regular solutions of the two equations baptized with his name. If we venture on an opinion we shall say that Nelson was the instigator of a promising revolution in the kingdom of quantum mechanics, but that he did not bring it to a successful issue.

## 5. Bibliography.